\documentclass[twocolumn,showpacs,preprintnumbers,aps,pra,amsmath,amssymb,floatfix,superscriptaddress]{revtex4}

\usepackage{graphicx}
\usepackage{dcolumn}
\usepackage{bm}
\usepackage{amssymb}
\usepackage[german, english]{babel}

\begin{document}

\title{Probing Ultracold Collisional Dynamics with Frequency-Chirped Pulses}

\author{M.J. Wright,$^{1,}$\footnote{Present address: Institut f$\selectlanguage{german}"u\selectlanguage{english}$r Experimentalphysik, Universit$\selectlanguage{german}"a\selectlanguage{english}$t Innsbruck, Technikerstra$\selectlanguage{german}"s\selectlanguage{english}$e 25, 6020 Innsbruck, Austria} J.A. Pechkis,$^{1}$ J.L. Carini,$^{1}$ and P.L. Gould}
\affiliation{Department of Physics, U-3046, University of Connecticut, Storrs, CT 06269}

\date{\today}

\begin{abstract}

	We report on the dynamics of ultracold collisions induced by near-resonant frequency-chirped light. A series of identical chirped pulses, separated by a variable delay, is applied to an ultracold sample of $^{85}$Rb, and the rate of inelastic trap-loss collisions is measured. For small detunings of the chirped light below the atomic resonance, we observe that the rate of collisions induced by a given pulse can be increased by the presence of an earlier pulse. We attribute this to the enhancement of short-range collisional flux by the long-range excitation of atom pairs to an attractive molecular potential. For larger detunings and short delays, we find that a leading pulse can suppress the rate of collisions caused by a following pulse. This is due to a depletion of short-range atom pairs by the earlier pulse. Comparison of our data to classical Monte-Carlo simulations of the collisions yields reasonable agreement.

\end{abstract}


\maketitle

	Recent years have witnessed enhanced capabilities in controlling both the external and internal degrees of freedom of atoms and molecules. Laser cooling and evaporative cooling of atoms \cite{Metcalf99} and the coherent control of excitation processes in molecules \cite{Rice00, Shapiro03} represent prime examples. The possibility of combining these two areas, i.e., applying coherent control techniques to ultracold systems, has generated a great deal of interest, especially in the context of using short laser pulses to produce ultracold molecules by photoassociating ultracold atoms \cite{Machholm94, Vardi97, Fatemi01, Vala01, Vatasescu01, Luc-Koenig04a, Luc-Koenig04b, Koch06a, Koch06b, Poschinger06, Salzmann06, Brown06}. Part of the appeal is the prospect of controlling the internal state distribution of the resulting molecules. Understanding and controlling the dynamics of the formation process will be key to these efforts. In the present work, we explore the nanosecond time-scale dynamics of a closely-related process: ultracold atomic collisions induced by frequency-chirped light \cite{Wright05}, shown in Fig. 1. By varying the delay between successive pulses of chirped light, we observe that the collisions induced by a given pulse can be either enhanced or suppressed by the presence of a preceding pulse, depending on the range of frequencies spanned by the chirp. If the chirp encompasses frequencies close to the atomic resonance, long-range excitation to the R$^{-3}$ potential (R is the internuclear separation) leads to collisional flux enhancement. For chirps centered well below the atomic resonance, efficient adiabatic excitation by the first chirp depletes the short-range atom pairs available to be excited by the second chirp. 
		
\begin{figure}
\centerline{\includegraphics{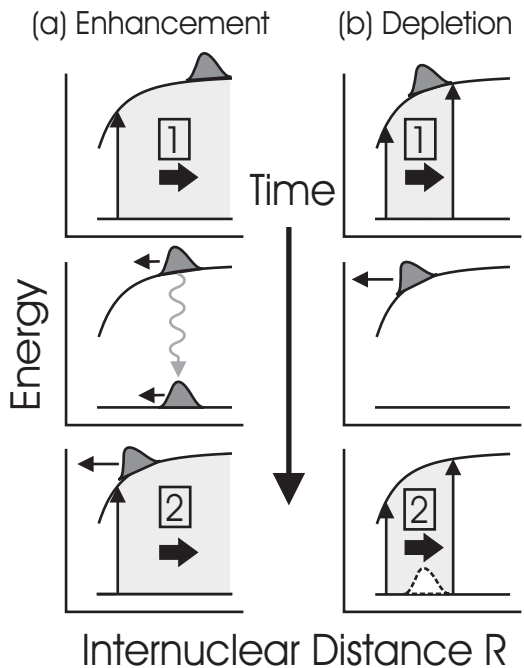}}
\caption{\label{fig:epsart} Schematic of the frequency-chirped excitation and resulting collisions, showing enhancement (a) and depletion (b) effects. Each frame shows the ground-state (5S+5S) and excited-state (5S+5P$_{3/2}$) potentials and the evolution of colliding atom pairs. The vertical lines delineate the range of frequencies spanned by each positive chirp (1 and 2). In (a), chirped pulse 1 excites atom pairs at long range (top frame). This range is centered close to the 5S $\rightarrow$ 5P$_{3/2}$ atomic resonance in (a) and well below resonance in (b). These pairs accelerate on the excited-state potential, then spontaneously decay back to the ground state (middle frame), providing enhanced short-range collisional flux for chirped pulse 2 (bottom frame). In (b), chirped pulse 1 excites atom pairs at short range (top). These excited pairs collide inelastically (middle), leaving a depleted pair distribution (dashed curve) to be sampled by chirped pulse 2 (bottom).}
\end{figure}

	A number of previous experiments \cite{Boesten96, Gensemer98, Orzel98, Fatemi01, Olivera03} have employed time-dependent excitation with fixed-frequency light to investigate dynamical effects in ultracold atomic interactions. The frequency-chirped light utilized in the present work is unique in two important ways \cite{Wright05}: 1) the chirp provides adiabatic, and therefore very efficient, excitation of atom pairs; and 2) the wide range of frequencies spanned by the chirp results in the nearly simultaneous excitation of atom pairs over a wide range of R.

	In the experiment, we illuminate trapped ultracold $^{85}$Rb atoms with pulses of frequency-chirped light and measure the resulting collisional rate constant $\beta$ for inelastic processes which lead to ejection from the magneto-optical trap (MOT) \cite{Wright05}. Such trap-loss collisions occur when atom pairs, initially excited to an attractive molecular potential by a chirped pulse, arrive at short range (e.g., R $<$ 100 a$_{0}$, where a$_{0}$ is the Bohr radius) in the excited state. The rate constant $\beta$ is determined by fitting the decay curve for the number of atoms in the MOT. There are two contributions to this decay: ultracold collisions occurring at a rate per atom $\beta$n, where n is the atomic density; and collisions with background gas occurring at a rate per atom $\gamma$. We are careful to operate at sufficiently low densities that radiative repulsion effects \cite{Walker90} are negligible, resulting in a constant effective volume for the MOT cloud. We also operate at low background pressures, $\sim$10$^{-10}$ torr, yielding long MOT lifetimes: $\gamma$${^{-1}}$$\sim$50 s. This is achieved by loading the primary MOT with a slow beam generated from a second MOT located in a separate vacuum chamber \cite{Lu96}. In addition to the collisions induced by the frequency-chirped light, the MOT itself has an inherent rate of trap-loss collisions. This contribution to $\beta$ is carefully measured by monitoring decays in the absence of the chirped light. The values for $\beta$ reported here have had this contribution subtracted. The chirped light can cause non-collisional perturbations to the MOT, especially when the chirp passes through resonance. The time-averaged fluorescence per atom and the volume of the MOT cloud can both increase (up to 3\% and 30\%, respectively) due to atomic excitation by the chirped light. These fractional changes are minimized by reducing the number of chirps per cycle, and are accounted for in determining absolute values of $\beta$. In the present work, we are interested in the collisional dynamics and, therefore, focus on the dependence of $\beta$ on the delay between successive chirped pulses.
	
	The primary MOT is configured in the phase-stable geometry \cite{Rauschenbeutel98} which reduces fluctuations in the properties of the trapped sample. It is operated with an axial field gradient of 12 G/cm, a total (sum of all six beams) peak intensity of 40 mW/cm$^{2}$, and a detuning of -1.5$\Gamma$ with respect to the 5S$_{1/2}$(F=3) $\rightarrow$ 5P$_{3/2}$(F'=4) cycling transition at 780 nm. Here $\Gamma$ = 2$\pi$(5.9 MHz) is the natural linewidth of this transition. A separate repumping laser, tuned to the 5S$_{1/2}$(F=2) $\rightarrow$ 5P$_{3/2}$(F'=3) transition, is used to prevent population from accumulating in the lower ground-state hyperfine level.
	
	The frequency-chirped light is produced by a rapid ramp of the current driving an external-cavity diode laser. To minimize the resulting amplitude modulation, a small portion of this chirped light is used to injection lock a separate ``slave'' diode laser \cite{Wright04}. This master-slave arrangement also results in a significantly higher output power being available to the experiment. The laser light is linearly polarized and focused onto the trapped sample. Its diameter ($\sim$100 $\mu$m) approximately matches that of the atom cloud. For the work reported here, the frequency increases linearly by 1 GHz in 100 ns, resulting in a chirp rate of +10 GHz/$\mu$s, and the peak intensity is fixed at 70 W/cm$^{2}$. An acousto-optical modulator (AOM) selects the central portion of each chirp, yielding a 40 ns FWHM Gaussian pulse.
	
\begin{figure}
\includegraphics{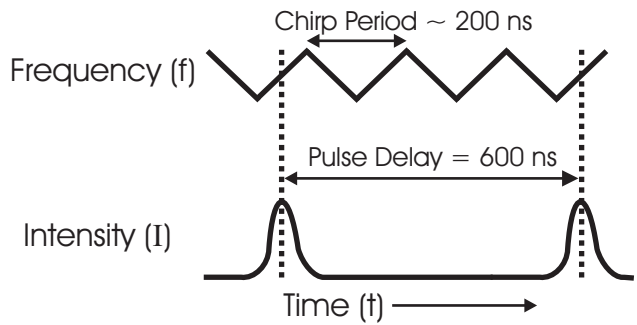}
\caption{\label{fig:epsart} Timing scheme for the multiple chirps. The laser frequency f undergoes a series of linear chirps. The same  portion of each chirp is selected by a synchronized acousto-optical modulator, resulting in a series of identical frequency-chirped pulses of intensity I. The delay between pulses is varied by selecting every n$^{th}$ pulse (n = 1, 2, 3, $\ldots$), thus maintaining the properties of the individual chirps. This figure, for example, shows n = 3.}
\end{figure}

	The timing of successive chirps is key to the present work and is shown in Fig. 2. The MOT is turned off for 150 $\mu$s every 722 $\mu$s. The repumping light remains on continuously in order to correct any optical pumping caused by the chirped pulses. During the MOT-off time, a train of chirped pulses, spaced by $\tau$, illuminates the MOT. A periodic current ramp is applied to the master laser, resulting in a symmetric triangle-wave frequency modulation (1 GHz amplitude, 200 ns period) of the slave laser output. The AOM selects the positive-slope portion of a given chirp cycle. The shortest possible delay between chirps ($\tau$=200 ns) is obtained by selecting adjacent positive ramps. Increasing delays are obtained by selecting every n$^{th}$ positive ramp. This ensures that the details of individual chirps do not change as the delay between chirps is varied. The number of chirps per MOT-off window is kept constant at either 40 (for the data in Fig. 3) or 80 (for the data in Fig. 4).
	
	We first examine the delay dependence of $\beta$ for a center detuning (relative to the 5S$_{1/2}$(F=3) $\rightarrow$ 5P$_{3/2}$(F'=4) cycling transition) of the chirp $\Delta_{c}$/(2$\pi$) = -300 MHz. In this case, the intensity of the chirped light is still rather high when it passes through the atomic resonance. Therefore, we expect significant excitation of atom pairs at long range (R$>$600 a$_{0}$). On the other hand, the chirp also encompasses larger detunings and the corresponding shorter-range excitations. We expect this combination of long-range and short-range excitations to lead to flux enhancement \cite{Sanchez-Villicana96}, as shown in Fig. 1a. In this process, a large number of atom pairs are initially excited at long range by light tuned near the atomic resonance. The atoms accelerate towards each other on the attractive potential, but because this curve is rather flat at long range, the atoms do not gain sufficient kinetic energy to escape from the trap before spontaneous emission returns the pair to the essentially flat ground-state potential. Although their kinetic energy is too low for escape, the atomic trajectories have been significantly altered. In particular, the atoms have been deflected towards each other and will therefore approach more closely than their original trajectories would have allowed. If the atom pair is now excited again, but this time at shorter range (by light detuned farther from the atomic resonance), the attractive molecular potential is much steeper and the excited atom pair will pick up sufficient energy to escape. The initial excitation at long range has thus enhanced the collisional flux available for the second excitation at short range. With fixed-frequency light (at both small and large detunings), this flux enhancement happens continuously \cite{Sanchez-Villicana96}. However, the trajectories themselves have a temporal dependence. We have previously probed these dynamics using delayed pulses \cite{Gensemer98}: a first pulse of near-resonant light to excite at long range, followed by second pulse of far-detuned light to re-excite at short range. When the delay between these pulses matches the time it takes the atoms to go from long range to short range, an enhancement in the collisional loss rate is seen. In the present work, the frequency chirp includes both the near-resonant and off-resonant light. The long-range excitation from one chirp enhances the flux available for short-range excitation by the following chirp. Since the excitation by the chirped light is time dependent, we expect the overall trap-loss collision rate to depend on the timing between successive chirps.
	
\begin{figure}
\centerline{\includegraphics{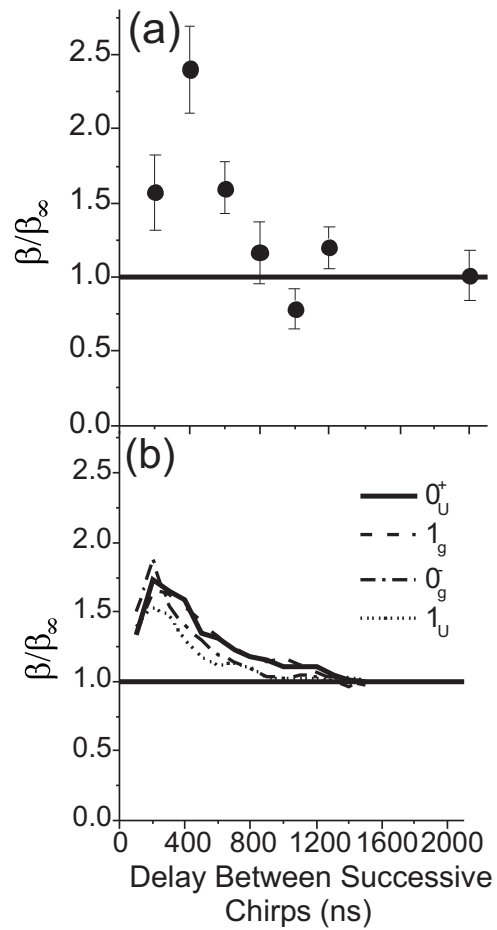}}
\caption{\label{fig:epsart} Dependence of collisional rate constant on delay between chirped pulses: (a) experimental results; (b) Monte-Carlo simulation results for different molecular states, as indicated. All results are normalized to the case of infinite delay. For the -300 MHz center detuning used here, enhancement is seen for short delays.}
\end{figure}

	In Fig. 3a, we show the collisional trap-loss rate constant $\beta$ as a function of delay between pulses of chirped light with $\Delta_{c}$/(2$\pi$)= -300 MHz. The results are normalized to the value at the longest delay (2 $\mu$s), since in this limit, the pulses act independently. The data do indeed display an enhancement peak centered at a delay of 400 ns. A maximum enhancement factor of 2.4$\pm$0.3 is observed. Results of Monte-Carlo simulations, discussed below, are shown in Fig. 3b. They show the same qualitative behavior as the data, but with a smaller peak occurring at shorter delay. We note that in the independent-pulse limit and using a time-averaged number of chirped pulses per second  $\nu_{c}$ = 5.5x10$^{4}$ s$^{-1}$, the absolute value of $\beta$ is 4.7x10$^{-12}$ cm$^{3}$s$^{-1}$. This result is slightly ($\sim$40$\%$) higher than, but consistent with, our previously measured value for these parameters \cite{Wright05} when we correct for $\nu_{c}$ and account for the factor of 2.0 enhancement (see Fig. 3a) resulting from the 500 ns delay used in that experiment. 
	
	We now examine the delay dependence for the case of a larger (more negative) center detuning for the chirp. A thorough discussion of the detuning dependence of the collision rate, for both positive and negative chirp directions, and at a fixed delay between chirped pulses, will be forthcoming \cite{Wright06}. For a center detuning of  $\Delta_{c}$/(2$\pi$)= -600 MHz, the frequency does not pass through the atomic resonance. Therefore, we expect that long-range excitation should not play an important role and that the flux enhancement discussed above should not occur. In fact, we expect just the opposite: depletion, as shown in Fig. 1b. Since the frequency-chirped light is rather intense, it is very efficient at adiabatically exciting atom pairs which have an internuclear separation R such that they are resonant at some point during the chirp. Immediately following the chirped pulse, all atom pairs within the spherical shell defined by the endpoints of the chirp will be excited via this rapid adiabatic passage. Since this excitation is primarily at short range, where the excited molecular potential is steep, the majority of these atom pairs will gain sufficient kinetic energy to escape from the trap. These pairs are thus not available to be excited by a second chirped pulse which immediately follows the first. We therefore expect fewer collisions to be induced by a second pulse when the delay is short. However, if this second chirped pulse is delayed sufficiently, the thermal atomic motion has time to fill in the depleted atom-pair distribution and the two pulses will act independently. We can estimate the expected persistence time for this depletion as $\Delta$R/v$_{t}$, where $\Delta$R$\sim$400 a$_{0}$ is the width of the spherical shell (from 500 a$_{0}$ to 900 a$_{0}$) and v$_{t}\sim$15 cm/s is the thermal velocity. This yields a time scale of $\sim$150 ns.
	
\begin{figure}
\centerline{\includegraphics{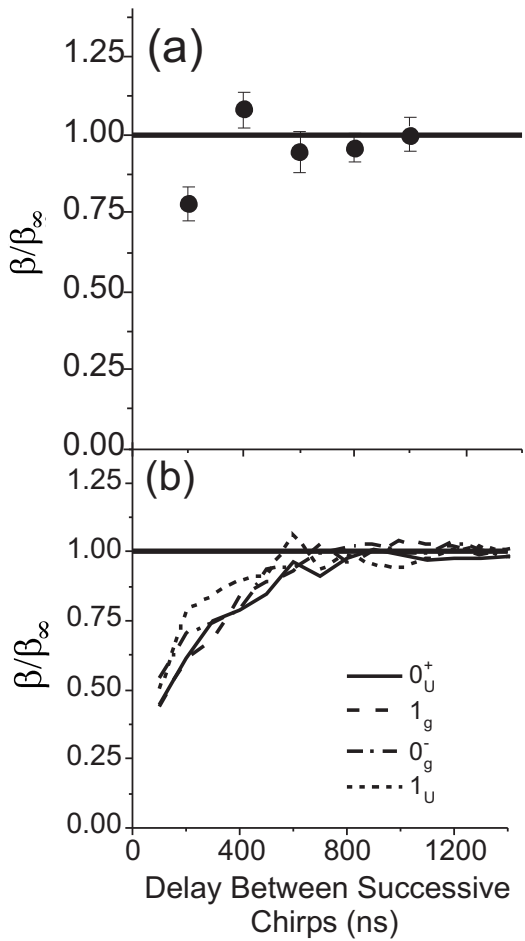}}
\caption{\label{fig:epsart} Dependence of collisional rate constant on delay between chirped pulses: (a) experimental results; (b) Monte-Carlo simulation results for different molecular states, as indicated. All results are normalized to the case of infinite delay. For the -600 MHz center detuning used here, depletion is seen at the shortest delay.}
\end{figure}

	In Fig. 4a, $\beta$ for this larger detuning of $\Delta_{c}$/(2$\pi$) = -600 MHz is plotted as a function of delay between successive pulses of chirped light. As in Fig. 3, the data are normalized to the long-delay (independent-pulse) limit. At the shortest achievable delay, 200 ns, a depletion of 22$\%\pm$6$\%$ is observed. The Monte-Carlo simulations, shown in Fig. 4b, are in reasonable agreement with the data. They predict an even larger depletion, $\sim$50\%, for delay of 100 ns, shorter than we have so far been able to achieve in the experiment.

	The simulated results shown in Figs. 3b and 4b are obtained from Monte-Carlo calculations of the collisions, which treat the atomic motion classically. The initial conditions for an atom pair, i.e., relative position and velocity vectors, are chosen randomly according to the appropriate distribution. A uniform spatial density and a temperature of 50 $\mu$K are assumed.  This pair is then subject to the first frequency-chirped pulse. The probability for excitation to an attractive molecular potential is then calculated using the Landau-Zener formula \cite{Landau32, Zener32}. A given atom pair is either excited or not, with a weighting given by this excitation probability. If the pair is excited, its motion on the attractive potential is followed until spontaneous emission (SE) occurs. The probability for SE is exponential with a time constant equal to the molecular state lifetime. Once SE occurs, the atom pair returns to the ground state, assumed to be flat, and follows a straight-line trajectory. Meanwhile, the second chirped pulse, delayed by $\tau$, illuminates the atom pair and the above calculation is repeated. If the atom pair arrives at sufficiently short range (R$<$100 a$_{0}$) in the excited state, as a result of excitation by either chirped pulse, that particular trajectory is considered a trap-loss event. A large number (e.g., 2x10$^{5}$) of trajectories are run and the fraction of them resulting in trap loss is a measure of relative value of $\beta$. This process is then repeated for various values of $\tau$ and for the four Hund's case (c) long-range molecular states: 0$_{u}^{+}$, 1$_{g}$, 0$_{g}^{-}$, and 1$_{u}$. The C$_{3}$ coefficients and lifetimes of these states are taken from \cite{Julienne91}. We do not include the 2$_{u}$ state in the simulations, because its decay (and excitation) is forbidden except at the very longest range \cite{Julienne91}.

	The simulations include only two frequency-chirped pulses whereas the experiment utilizes a train of N$>>$1 (typically N=40 or 80) pulses per MOT-off window. We can consider two contributions to $\beta$($\tau$): $\beta_{1}$ from a single chirp and $\beta_{2}$ arising from the interaction between two successive chirps. Note that $\beta_{1}$ is independent of $\tau$ (for $\tau$=$\infty$, the simulation result is 2$\beta_{1}$), while $\beta_{2}$ vanishes for $\tau$=$\infty$. The value of $\beta_{2}$ can be either positive, indicating enhancement, or negative, indicating depletion. From the $\tau$ dependence of the simulated value of $\beta$, we can extract $\beta_{1}$ and $\beta_{2}$. The quantity to be compared with experiment would then be N($\beta_{1}$+$\beta_{2}$). Since we are interested primarily in the dependence on delay, the comparisons are made using values of $\beta$ normalized to the case of infinite delay.

	As seen in Figs. 3 and 4, the simulated values of $\beta$ describe the experimental results rather well. It is interesting that the different molecular potentials all exhibit a similar delay dependence, despite the fact that their radiative lifetimes and C$_{3}$ coefficients vary significantly. For the -300 MHz center detuning (Fig. 3), the enhancement peak in the simulation occurs at a somewhat shorter delay and is less pronounced in comparison to the experiment. For the -600 MHz center detuning (Fig. 4), the depletion is more pronounced in the simulation than in the experiment. These differences may be due to simplifications in the simulations. The motion is treated classically and the effects of hyperfine structure are not included. Also, in a given run of the simulation, a single excited molecular potential is assumed.

	In summary, we have investigated the dynamics of ultracold atomic collisions induced by pulses of frequency-chirped light. We find generally good agreement between our measurements and the results of Monte-Carlo simulations. The rate of inelastic trap-loss collisions caused by a given chirped pulse is modified by the presence of a preceding pulse. The extent of this modification depends on the delay between the two pulses. Varying this delay allows us to probe the collisional dynamics. We see two main effects. First, when the chirp includes frequencies near the atomic resonance, the resulting excitation of long-range atom pairs by the first pulse leads to an enhancement of the collisional flux available for the second pulse. The time scale for this enhancement is set by the trajectories of excited atom pairs which have decayed at long range. Second, when the chirp includes frequencies far from resonance, short-range atom pairs are efficiently excited and caused to collide by the first pulse, leading to a reduction in collisions induced by the second pulse. This depleted distribution of short-range atom pairs is eventually filled in by the thermal motion. The fact that we see significant depletion at the shortest delay indicates that the chirped excitation of available atom pairs by rapid adiabatic passage is indeed efficient.

	Our findings have several important consequences that are relevant to efforts to form ultracold molecules by pulsed photoassociation of ultracold atoms. The time-dependent enhancement in collisional flux caused by chirped-pulse excitation of long-range atom pairs may benefit these efforts. On the other hand, our observation of depletion indicates that this process must be considered in experiments utilizing high-repetition-rate short-pulse lasers. The cold atoms fill in the depleted pair distribution rather slowly. Finally, we point out that our depletion effect is closely related to the production of a ``hole'' in the ground-state collisional wavefunction \cite{Luc-Koenig04a, Koch06b}. Such a hole is associated with mixing bound levels into the continuum state and is predicted to lead directly to the production of weakly-bound ground-state molecules.

	We thank Steve Gensemer and Christiane Koch for useful discussions. This work was supported in part by the Chemical Sciences, Geosciences and Biosciences Division, Office of Basic Energy Sciences, U.S. Department of Energy.

\end{document}